\documentclass[smallextended]{svjour3}
\usepackage{amssymb}
\usepackage{amsmath}
\usepackage{graphicx}
\usepackage{txfonts}
\usepackage{booktabs}
\usepackage{textcomp}
\usepackage{color}
\usepackage[caption=false]{subfig}
\usepackage{url}
\usepackage{xcolor}
\RequirePackage{fix-cm}

\usepackage[utf8]{inputenc}
\begin{document}

\title{A robotic pipeline for fast GRB followup with the Las Cumbr\'es Observatory Network}
\titlerunning{Pipeline paper}
\authorrunning{R.Martone et al.}

\author{R.~Martone 
           \and
           C.~Guidorzi
           \and 
           C.~G.~Mundell
           \and
           S.~Kobayashi
           \and
           A.~Cucchiara
           \and
           A.~Gomboc
           \and
           N.~Jordana
           \and
           T.~Laskar
           \and
           M.~Marongiu
           \and
           D.~C.~Morris
           \and
           R.~J.~Smith
           \and
           I.~A.~Steele
           }

\institute{R.~Martone (mrtrnt@unife.it) \and C.~Guidorzi \and M.~Marongiu \at Department of Physics and Earth Science, University of Ferrara, via Saragat 1, I--44122, Ferrara, Italy\\
\and R.~Martone \and M.~Marongiu \at ICRANet, Piazzale della Repubblica 10, I--65122, Pescara, Italy\\
\and C.~G.~Mundell \and  N.~Jordana \and T.~Laskar \at Department of Physics, University of Bath, Claverton Down, Bath, BA2 7AY, UK\\
\and S.~Kobayashi \and R.~J.~Smith \and I.~A.~Steele \at Astrophysics Research Institute, Liverpool John Moores University, Liverpool, L3 5RF, UK\\
\and A.~Cucchiara \and D.~C.~Morris \at University of the Virgin Islands, \#2 John Brewers Bay, 00802 St Thomas, VI, USA\\
\and A.~Gomboc \at Centre for Astrophysics and Cosmology, University of Nova Gorica, Vipavska 11c, Ajdov\v s\v cina 5270, Slovenia
    }
   \date{}

\abstract
  {In the era of multi-messenger astronomy the exploration of the early emission from transients is key for understanding the encoded physics. At the same time,  current generation networks of fully-robotic telescopes provide new opportunities in terms of fast followup and sky coverage.}
  {This work describes our pipeline designed for robotic optical followup of gamma-ray bursts with the Las Cumbr\'es Observatory
   network.}
 {We designed a Python code to promptly submit observation requests to the Las Cumbr\'es Observatory network within 3 minutes of the receipt of the socket notice. Via Telegram the pipeline keeps the users informed, allowing them to take control upon request.}
{Our group was able to track the early phases of the evolution of the optical output from gamma-ray bursts with a fully-robotic procedure and here we report the case of GRB180720B as an example.}
{The developed pipeline represent a key ingredient for any reliable and rapid (minutes timescale) robotic telescope system. While successfully utilized and adapted for LCO, it can also be adapted to any other robotic facilities.}

\maketitle

\section{Introduction}
\label{sec:intro}
Gamma-ray bursts (GRBs) are detected as short-lived, intense flashes of high-energy $\gamma$-rays \cite{Klebesadel73} so-called prompt emission, produced by the core-collapse of a massive star (see \cite{Woosley12} for a review) or the merger of two neutron stars \cite{Lipunov95,Abbott17} or a neutron star and stellar-mass black hole \cite{Eichler89,Paczynski91,Narayan92}. Propagation of the expanding fireball into the circumburst medium produces long-lived radiation - the afterglow - that is detectable across the electromagnetic spectrum; detection of the first optical afterglow \cite{Vanparadijs97} confirmed the cosmological origin of GRBs and highlighted the need for accurate prompt localisations and increasingly rapid followup observations. The first prompt optical emission was detected by \cite{Akerlof99} in 1999, who discovered the possibility to study just the outburst to undestand its processes. The launch of NASA's Neil Gehrels {\em Swift} in 2004 \cite{Gehrels04} opened a new era of rapid-response multimessenger astronomy,  providing unprecedented real-time discoveries of GRBs with arcmin localisations of prompt $\gamma$-ray emission. It's became clear to develope the global  network of fully robotic telescopes, to have 24h of night time in both hemispheras (ROTSE \cite{Akerlof03}, MASTER \cite{Lipunov10}, TAROT \cite{Boer99},  BOOTES \cite{Castro-Tirado99}). In parallel, the development of the world's largest fully autonomous robotic optical telescopes such as the 2-m  Liverpool Telescope \cite{Steele04} and the identical 2-m Faulkes Telescopes \cite{FTpaper1,FTpaper2} that can respond to GRB discoveries within minutes of the alert notification provided new insights into the nature of the early afterglow \cite{Monfardini06,Mundell07,Melandri08,Virgili13,Kopac15b}, the physics of reverse shocks \cite{Gomboc08,Gomboc09,Japelj14,Laskar16,Laskar18}; and the importance of ordered magnetic fields in the relativistic ejecta \cite{Mundell07sci,Steele09,Mundell13,Troja17nat}. The rapid evolution of the blast-wave emission and the complexity of early time light curves within the first few minutes to hours of the GRB drove the development of autonomous software systems for immediate response to the GRB trigger combined with rapid, automatic identification, classification and selection of subsequent followup observations \cite{Guidorzi06,Mundell10}.  

The development of Las Cumbr\'es Observatory that now includes the two 2-m Faulkes Telescopes in a network with nine 1-m and nine 0.4-m telescopes has opened new opportunities for global monitoring of time-variable and transient phenomena at optical and infrared wavelengths. Here we present a new GRB pipeline that builds on our experience of developing autonomous systems for immediate GRB followup \cite{Guidorzi06}. The goal of this pipeline is to optimise receipt of and response to transient triggers --  particularly GRBs -- for which the location and time of discovery is completely unpredictable and for which a rapid response is vital to maximise the chance of capturing the physics encoded in the prompt optical and early afterglow radiation. This is to ultimately determine the nature of the progenitors, energetics and explosion physics. Whilst optimised for GRB followup, the pipeline is of general use for other kinds of multi-messenger trigger e.g. gravitational wave, neutrino, high-energy cosmic rays, supernovae (SNe) and could be extended to other telescope networks.  

\section{The Las Cumbr\'es Observatory}
\label{sec:LCO}
\subsection{The Network}

\begin{table*}
\centering
\caption{LCO telescopes and their localization.}
\label{tab:observatories}
\begin{tabular}{cccc}
\hline
Emisphere & Name & Location & Code \\
\hline
North & McDonald Observatory & Texas-USA & ELP \\
      & Haleakala Observatory & Hawaii-USA & OGG \\
      & Wise Observatory & Israel & TLV \\
      & Teide Observatory & Canary Islands-Spain & TFN \\
\hline
South & Cerro Tololo Inter-American Observatory & Chile & LSC \\
      & Siding Spring Observatory & Australia & COJ \\
      & South African Astronomical Observatory & South Africa & CPT \\
\hline
\end{tabular}
\end{table*}

The Las Cumbr\'es Observatory Global Telescope Network\footnote{\url{https://lco.global}} \cite{LCO13} is a network designed for time domain astronomy at optical and near-IR wavelengths. It consists of a series of telescopes located at different sites in both the northern and southern hemispheres and whose schedules are managed by a common scheduler. The presence of this central ``brain'' makes the network able to work as a single facility, with the obvious advantages of many observing points.

The network is owned and operated by the Las Cumbr\'es Observatory Global Telescope (LCOGT), a young organisation whose main goals are providing professional scientists affordable and reliable instruments to get professional-class data, and non-professional users an easy and user-friendly way to feed their interest with professional instruments in a supervised environment.

The LCOGT organisation pursues this mission using a set of small to medium-aperture instruments spread among four continents: the three classes of $0.4$, 1, and 2 meters are available at the present moment. The basic developing strategy requires the presence of all this three main classes of telescopes in each site: the idea is to have medium-class instruments to study faint objects and something able to lighten their load  when their larger aperture is not needed: a 0.4-meters telescope is perfect to handle this task. The available sites, their location and international identification code can be found in Table~\ref{tab:observatories}, while additional information about on-site instrumentation can be found in \cite{LCO13}. Current-state information can be found at the LCO web site.

\begin{table*}
\centering
\caption{Table of past and present proposals by our group awarded with time on LCO network.}
\label{tab:proposals}
\begin{tabular}{ccccc}
\hline
Name & Title & PI & start & end \\
\hline
CON2014A-001 & “Fast Transients in the Era of Rapid Followup” & C. Mundell & 01-01-14 & 31-12-14\\
ARI2015A-001 & “Studies of Gamma-ray Bursts and the associated Supernovae” & S.
Kobayashi & 01-01-15 & 31-12-15\\
ARI2016A-004 & “Studies of Gamma-ray Bursts and the associated Supernovae” & S.
Kobayashi & 01-01-16 & 31-12-16\\
ARI2017AB-005 & “Followup observations of GRBs” & S.
Kobayashi & 01-01-17 & 31-12-17\\
ARI2018A-002 & “Followup observations of GRBs” & S.
Kobayashi & 01-12-17 & 30-06-18\\
 ARI2018B-001 & “Followup observations of GRBs” & S.
Kobayashi & 01-06-18 & 30-11-18\\
\hline

\hline
\end{tabular}
\end{table*}

\subsection{Instruments}
\label{subsec:instruments}
\begin{table*}
\centering
\caption{Telescopes main specs.}
\label{Telescopes}
\begin{tabular}{ccccc}
\hline
Aperture & f/$\#$ & Mount & slew rate & Location \\
m &  &  & degrees $s^{-1}$ & \\
\hline
1 & f/10 & Alt/Az & 2 & [COJ, OGG] \\
\hline
2 & f/8 & Equatorial & 6 & [COJ, CPT, LSC, ELP] \\

\hline
\end{tabular}
\end{table*}

\begin{table*}
\centering
\caption{Instruments main specs.}
\label{Instruments}
\begin{tabular}{ccccccc}
\hline
Name & Detector format & Readout & Filters & Mounted on (class) & FOV & Pixel size \\
 & ($\mu m$) & (s) & & (m) & (') & ('')\\
\hline
Sinistro & 4096x4097x15 & 4 & UBVRI, u’ g’ r’ i’ z’ sYw & 1 & 26.5 & 0.389\\
\hline
Spectral & 4096x4097x15 & 14 & UBVRI, u’ g’ r’ i’ z’ sYw, $H_{alpha}$,  & 2 & 10.5 & 0.304\\
 & & & $H_{beta}$, OIII, D51, UV, v & & & \\
\hline

\hline
\end{tabular}
\end{table*}

The LCO network gives easy access to professional-class instruments for both imaging and spectroscopy on all the cited classes of telescopes, depending on the sites. Our past and present proposals (see Table ~\ref{tab:proposals}) are focused  on imaging with only 1 and 2 meters telescopes, so we briefly describe these classes together with the imaging instruments mounted on them.

The 2-m telescopes are present in:
\begin{itemize}
\item{COJ - 1 unit}\footnote{Faulkes Telescope South (FTS)}
\item{OGG - 1 unit}\footnote{Faulkes Telescope North (FTN)}
\end{itemize}
These are two Faulkes Telescopes featuring a Ritchev-Chr\'etien Cassegrain f/10 optics and an Alt-Az configuration. These telescopes are capable to perform rapid repointing with a turnaround of 45 seconds on average, thanks to a maximum slewing speed of 2 degrees/s. These details are summarised in Table~\ref{Telescopes}. The imaging instrument available on this class is the "Spectral"\footnote{\url{https://lco.global/observatory/instruments/spectral/}} and the FLOYD spectrograph\footnote{\url{https://lco.global/observatory/instruments/floyds/}}.

The 1-m telescopes are available in:
\begin{itemize}
\item{COJ - 2 units}
\item{CPT - 2 units}
\item{LSC - 3 units}
\item{ELP - 2 units}
\end{itemize}
These telescopes feature a Ritchev-Chr\'etien Cassegrain f/8 optics and an Alt-Az configuration. This class is capable to perform a very rapid repointing with a turnaround of less than 30 seconds, thanks to a slewing speed of 6 deg/s. These details are summarised in Table~\ref{Telescopes}. This class features the ``Sinistro'' imaging instrument\footnote{\url{https://lco.global/observatory/instruments/sinistro/}}and a the NRES spectrograph\footnote{\url{https://lco.global/observatory/instruments/nres/}}.

All the observation sites are equipped with a local weather station, providing information on humidity levels, weather conditions, wind speed, temperature, sky transparency, and other relevant quantities. This information, together with the operational statuses of the instruments, are used by the automatic system that controls the aperture of the dome and that communicates with the scheduler. The same info is available to users and can be queried both via the web interface\footnote{\url{https://weather.lco.global/#/bpl} (weather)

\url{https://lco.global/observatory/status/} (instruments status)} or through the central database via API scripts (Sect.~\ref{subsec:visibility}).

\subsection{Observation requests handling and management}
A global network like the LCOGT requires a very high level of automation, so all the steps from the handling of observation requests to observations are managed from the central headquarter located near Santa Barbara, California and automatically executed by the on-site node.
LCOGT network entered in an agreement with the NOAO and other partners offering observing time as part of their general call for proposals.

Observations requests can be submitted both via the web interface\footnote{\url{https://observe.lco.global/create/}} or through API scripts\footnote{\url{https://developers.lco.global/?python#introduction}}.
A request consists of a certain number of exposures, for which the required instrumental configuration (e.g., aperture, filters, exposure times) must be specified. Further constraints (e.g., seeing, transparency, moon distance) can be also set. The system provides also the possibility to push for high priority request, through a protocol called ``Target of Opportunity'' (ToO): this is designed for observations of transients and relatively short-lived phenomena. This kind of requests is granted a priority route to go quickly through all the different steps up to the final node. The ToO requests can override previous, lower-priority observations, that will be resumed at its end.

Once the request is complete and submitted, the scheduler funnels it through a specific node on the basis of the user requests and of the information from the node itself, such as visibility of the target, telescopes availability, weather conditions, and other relevant quantities.

When individual exposures are completed, the relative temporary-reduced, quick-look files are produced and made available on the proposal web page along with the original raw files. The former ones are produced from the latter ones using a fast, light version of the data pipeline described in \cite{LCO13}, while the finally reduced products are usually released several hours later using the complete, slower version of the same pipeline. All these products can be downloaded both via the web interface or through API scripts\footnote{\url{https://developers.lco.global/#get-related-frames}}.

\subsection{Our GRB-driven proposals}
\label{subsec:proposal}
The GRB pipeline described in this paper has been developed under proposal ARI2018B--001 (PI: S. Kobayashi), whose aim is monitoring of GRB/SNe optical light curves (Sect.~\ref{sec:intro}).
The main goal is the prompt followup of GRB discovered by space-based observatories like the Neil Gehrels {\em Swift} Observatory or {\em INTEGRAL} \cite{Lavigne98}. The development phase has exploited the great experience our group has gained since our first observations with the LCO network in 2014: a brief history of our proposals on the LCO network is reported in Table~\ref{tab:proposals}.
The proposal on which we developed our code, ARI2018B--001, has been granted the following observation time:
\begin{itemize}
\item{3 hours at standard priority on 2-m class units;}
\item{$6.5$ hours at high priority (ToO) on 2-m class units;}
\item{7 hours at high priority (ToO) on 1-m class units.}
\end{itemize}
The observing period began June 1, 2018, and lasted for six months.


\section{GRB pipeline description}
\label{sec:strategy}
\subsection{GCN analysis}
\label{subsec:GCN}
The GRB pipeline\footnote{\url{https://github.com/kobe90/LCO_pipeline}} is designed to optimally exploit the potential of the LCO network, in particular: (i) its worldwide distribution of observing facilities, that allows a good sky coverage in both hemispheres; (ii) prompt response to ToO requests. It continuously listens to the Gamma-ray burst Circular Network (GCN \cite{Barthelmy95}) Notices\footnote{The system of Notices and Circulars is described at \url{https://gcn.gsfc.nasa.gov}} and, in case of a promptly observable target, reacts by promptly submitting a customised observation sequence. A rapid and effective management of this phase is essential to track the early-time light curve of the possible optical afterglow: in this respect, reducing at minimum the human-in-the loop involvement is critical to avoid possible mistakes and rapidly perform the most appropriate observing sequence.
 
The first task is the continuous monitoring of the trigger notices from both  {\em Swift} and {\em INTEGRAL} missions. Every such notice is a message distributed via socket and e--mail that provides basic information about a trigger from one of satellites connected to the Inter--Planetary Network (IPN)\footnote{\url{https://heasarc.gsfc.nasa.gov/w3browse/all/ipngrb.html}}, such as the time of the event, the detector, the kind of trigger and other secondary parameters. In case of an event that the on-board system recognises to be a GRB, an estimation of the position is provided too, together with the related uncertainty.

We conveniently make use of the socket-distributed version of the GCN notices, that consists of a text message formatted in the JavaScript Object Notation (JSON)\footnote{\url{https://www.json.org}}, a light format that is both human and machine readable.

To ensure the maximum portability and a smooth interaction with the LCO network, the whole pipeline presented in this paper is written in Python~$3.6$. The code has been tested both under MacOs~$10.12$ and on Fedora~21.

The handling of incoming, socket-distributed notices is performed using the \texttt{gcn} Python library\footnote{\url{https://github.com/lpsinger/pygcn}}. The basic structure of this part of the code is composed by a client that continuously waits for notices distributed via the VOEvent Transport Protocol\footnote{\url{http://www.ivoa.net/documents/Notes/VOEventTransport/}}. When a VOEvent is received, the socket notice is passed on to a handler, that extracts the required information. Among the setting options, the user can easily select the type of events that will set the handler on.

The GCN-Notice system can generate different kinds of VOEvent, that are reported at a web page\footnote{\url{https://gcn.gsfc.nasa.gov/filtering.html}.} and that correspond to different actions and alerts.
As of September 2018, the GRB pipeline is designed to handle {\em Swift} and {\em INTEGRAL} Burst Notices, so we used the so-called "BAT Position Notice"\footnote{\url{https://gcn.gsfc.nasa.gov/swift.html}} (for {\em Swift}) and "REFINED Notice" (for {\em INTEGRAL})\footnote{\url{https://gcn.gsfc.nasa.gov/integral.html}}. 

Once an event of the selected type is received, the handler can extract information using the \texttt{xml} syntax. We extract the following basic pieces of information:
\begin{itemize}
\item{Name of the trigger (which uniquely identifies it);}
\item{Position: right ascension (RA) and declination (DEC);}
\item{Position error radius;}
\item{galactic latitude.}
\end{itemize}
The VOEvent is further checked to see whether it matches our interests by looking at the boolean values of the following variables:
\begin{itemize}
\item{not-a-GRB variable: if \texttt{False} the event is probably a real GRB;}
\item{catalog variables: if at least one is \texttt{True}, the emission that triggered the instrument was probably from a known source, i.e. uninteresting for a GRB programme;}
\item{test variable: if set to \texttt{True} the VOEvent was the result of a test and not a real event.}
\end{itemize}
The extracted position is then double checked against our local catalogue of known variable X-ray sources that can occasionally mimic the occurrence of a GRB.

Notices from {\em INTEGRAL} are slightly different, in that they lack the catalogue variable, only the information contained in the first (not-a-GRB) and third (test variable) labels is processed. Also in this case the position of the burst candidate is crosschecked against our local catalogue.

\begin{figure}
\centering
\subfloat[visibility-north]{
\includegraphics[clip,width=\columnwidth]{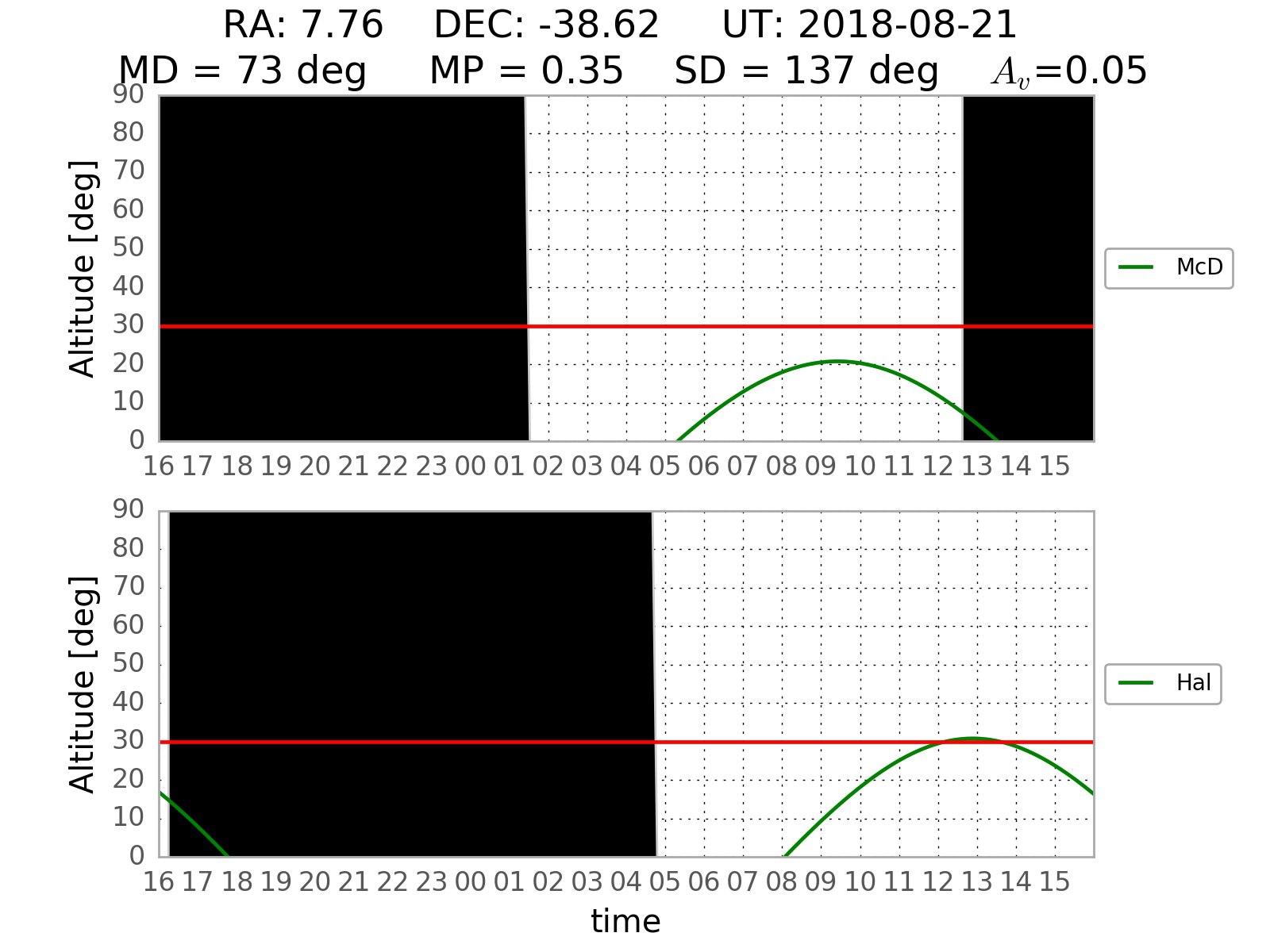}
}

\subfloat[visibility-south]{
\includegraphics[clip,width=\columnwidth]{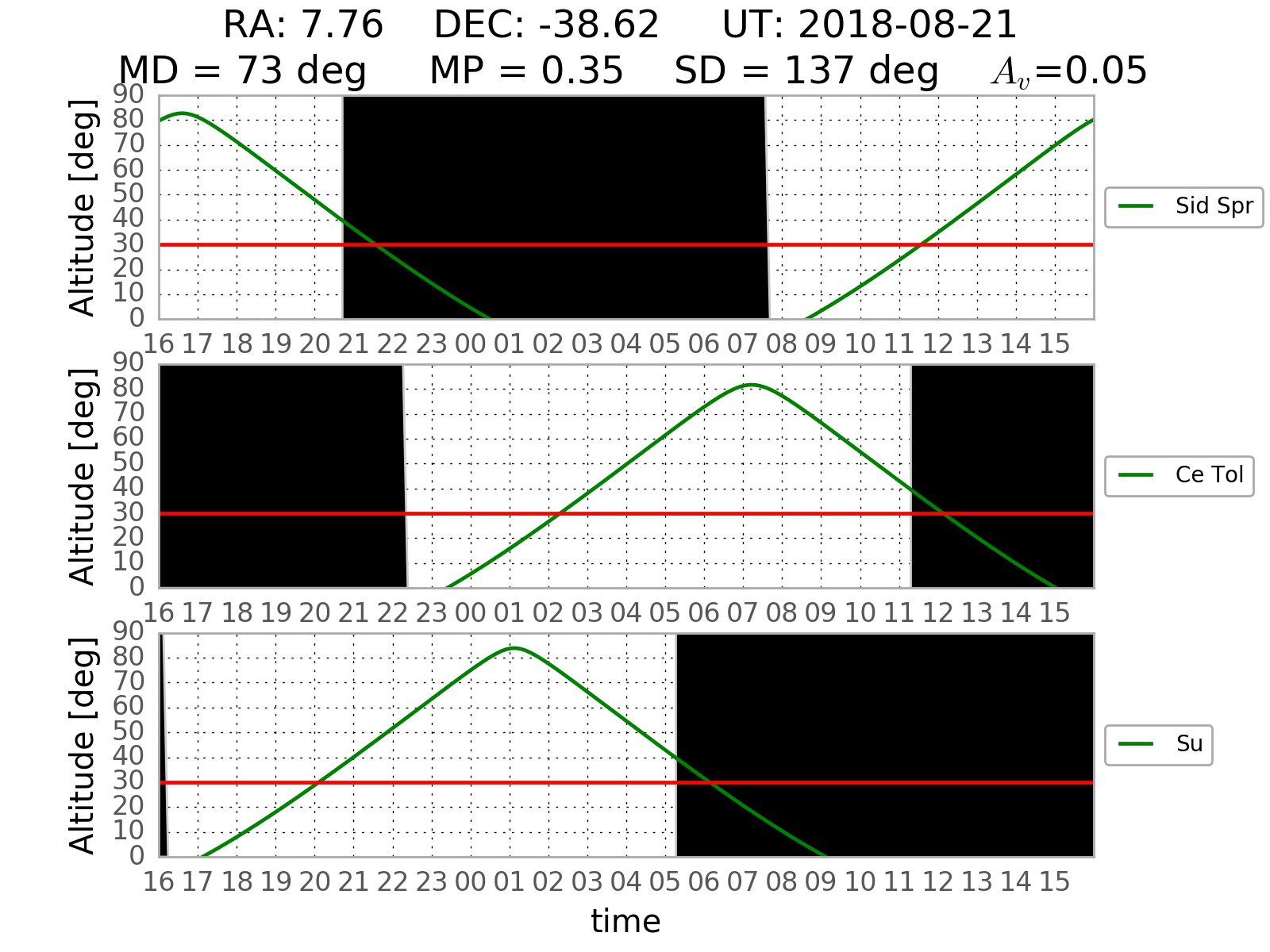}
}
\caption{Visibility plot (altitude vs Universal Time (UT)) for the northern (a) and southern (b) hemisphere sites of the LCO network we access to. The black regions indicate the daytime, while the red, horizontal line indicates the minimum altitude at which the telescopes can observe. On the right-hand side the observatory shortened names are indicated, while on the top further burst information is summarised: equatorial coordinates (J2000), UT, MD (Moon Distance), MP (Moon Phase), SD (Sun Distance), $A_{v}$ (Galactic visual extinction).}
\label{visibility_plot}
\end{figure}

The same information is also used to build two visibility plots (one per hemisphere) for the LCO sites accessible to our proposal. An example of this kind of products can be found in Figure~\ref{visibility_plot}. Sun distance, Moon distance and phase, together with the Galactic visual extinction, are also calculated. The last is estimated using the SFD radio maps \cite{Schlegel98}, that can be easily queried using the \texttt{SFDmap} library\footnote{\url{https://github.com/kbarbary/sfdmap}}. Only candidates for which the expected visual extinction is $A_V<5$~mag are considered. Note that the described visibility plots are built using only the trigger and the observatory locations, regardless of the actual availability and status of each telescope.

\subsection{Visibility of the target}
\label{subsec:visibility}
As briefly outlined in Section~\ref{subsec:instruments}, the availability of a certain telescope can be queried via a boolean output (answering the question ``Can the telescope work now?'') and with a certain number of input arguments that can be grouped into two main categories: instrumental status and weather conditions. These pieces of information have to be combined with that contained in the visibility plots to know the effective visibility of the event from each instrument. A certain target is qualified as observable by a certain telescope if:
\begin{enumerate}
\item{its altitude is $\ge 30^\circ$;}
\item{the Sun is $>12^\circ$ below horizon (Nautical Dawn/Sunset);}
\item{the Moon distance is $\ge 30^\circ$;}
\item{good local weather conditions;}
\item{the telescope status is operational.}
\end{enumerate}
Requirements 1-3 can be answered from the burst and telescope positions, while 4 and 5 can be easily checked querying the LCO network.
We used some templates \footnote{\url{https://developers.lco.global/?python#retrieve-weather}.} to build and test a function that checks data collected by on-site weather sensors. The environmental conditions are considered good, whenever all the following requirements are fulfilled:
\begin{itemize}
\item{air temperature $>-20$~$^\circ$C;}
\item{air humidity $<90$~\%;}
\item{wind speed $<15$~km/h;}
\item{sky brightness $<18$~mag/arcsec$^2$}.
\end{itemize}

Finally, the operational status of each telescope is checked through a function that returns the list of available units\footnote{examples are available at \url{https://developers.lco.global/?python#telescope-states}.}).

The list of instruments that can effectively observe the target is then stored for the submission phase and is also provided to users in a compact format, as described in the following.

\subsection{Distribution to the users}
\label{subsec:distribution}
A user is anyone that receives the GRB pipeline notices through one of the distribution channels (described in the following), and that is enabled to interact and take control of the followup activities.

As far as the choice for the optimal strategy is concerned, human action is preferable. Upon receiving essential, readable-on-the-fly information, any user aiming at taking control, has to explicitly reply to the GRB pipeline. In this case, the pipeline sends a confirmation of the user choice (both to him and other users) and then comes back to the listening mode.

The promptly distributed information includes:
\begin{itemize}
\item{visibility plots;}
\item{list of instruments able to promptly observe the event and their location;}
\item{amount of residual observing time of our proposal.}
\end{itemize}

While the first two items are already described above in Section \ref{subsec:visibility}, the last one is obtained querying the LCO proposal repository\footnote{Some templates are available:  \url{https://developers.lco.global/?python#proposals}.}. The observing time is classified into categories based on the instrument class (1m or 2m in the case of our proposal) and on the observing priority: for obvious reasons the GRB pipeline only sends ToO requests.

Communications with users is possible through two channels: (i) via ordinary emails; (ii) via Telegram\footnote{\url{https://telegram.org}}, a multi-platform open-API messaging system with bidirectional communication enabled. While the interaction with mail servers can be done routinely in Python using a wide collection of specific libraries, we opted for using the \texttt{Python-telegram-bot} library\footnote{\url{https://github.com/python-telegram-bot/python-telegram-bot}} to manage Telegram messages both ways. The Telegram channel makes use of a bot that acts as an ordinary, human user, distributing messages.
A person that is interested in our Telegram notice just needs to create a Telegram account and contact our bot: this way they will be added to our list of interacting users. 
Unlike the Telegram notice, the email is a sort of reminder and lacks the possibility of interacting with the pipeline: the idea is that very rapid, interacting Telegram notices are dedicated to a smaller group of core users strongly involved and more interested in taking control whenever this is considered as a preferable option.

\begin{figure}
\centering
\includegraphics[width=9cm]{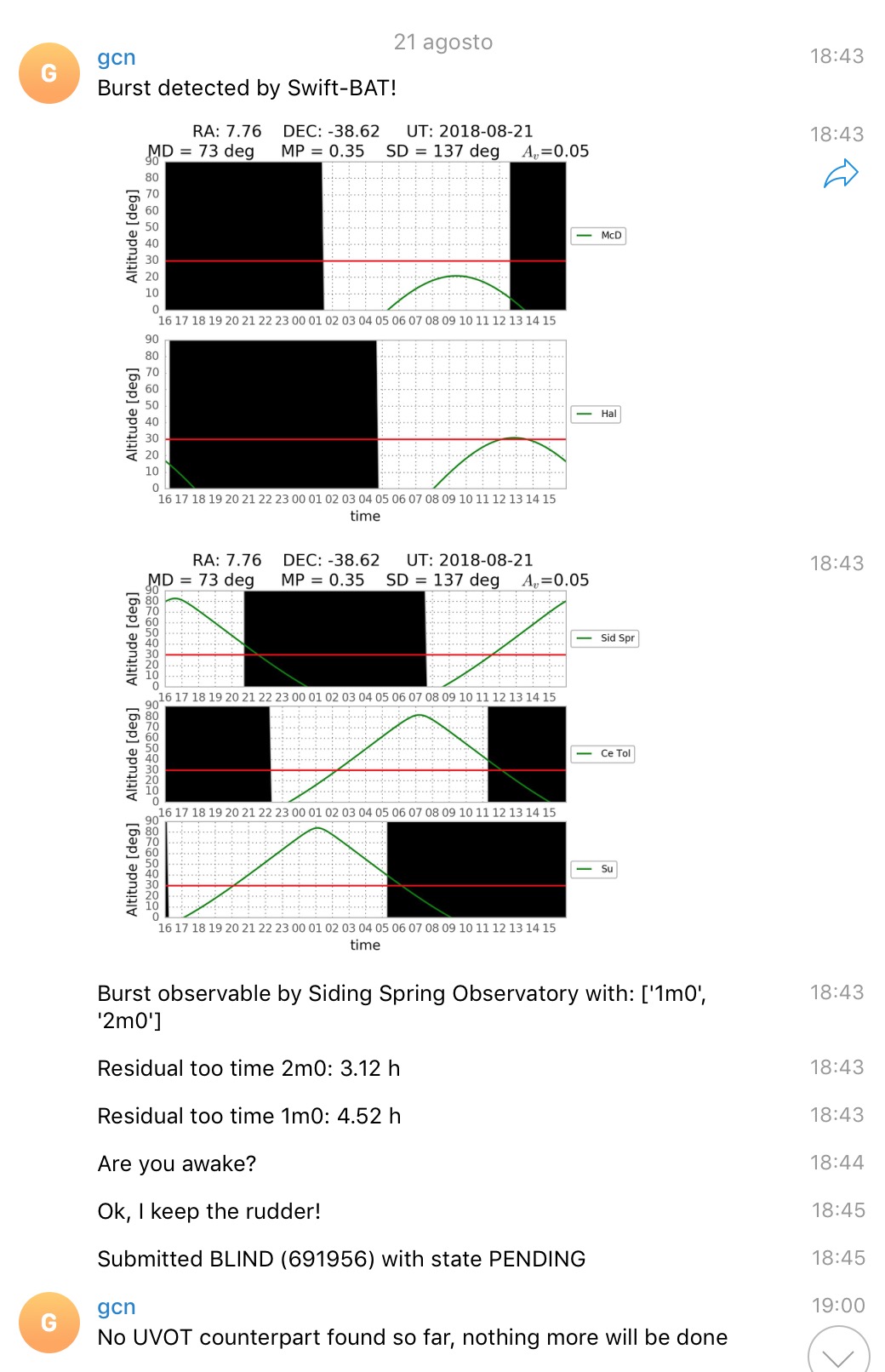}
\caption{Example of a notice distributed via Telegram. In this case the users remain quiet, thus letting the pipeline move on to the submission phase.}
\label{Telegram_obs_ex}
\end{figure}

\begin{figure}
\centering
\includegraphics[width=9cm]{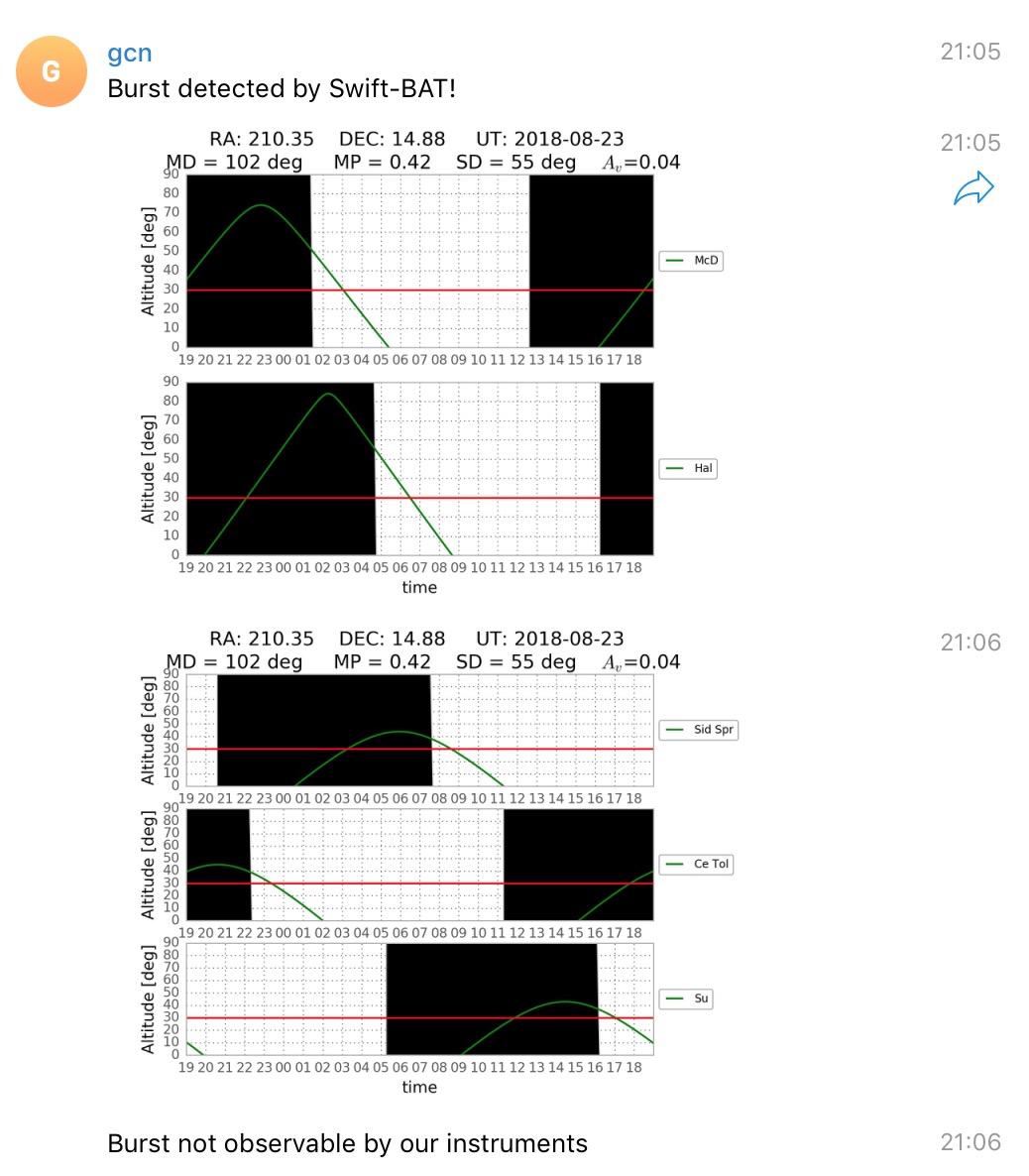}
\caption{Example of a notice distributed via Telegram in case of non-observable target.}
\label{Telegram_not_obs_ex}
\end{figure}

An example of Telegram interaction is provided in Figure~\ref{Telegram_obs_ex}.
After the ``Are you awake?'' question the programme waits for an answer from any of the users for one minute; if no one replies, it autonomously submits the observation request (Section~\ref{subsec:submission and getting products} for further details). Differently, the user that responds  is notified to be in charge (while other users are notified that someone else is in charge). Consequently, the pipeline stops taking care of the present trigger and gets back to the GCN Notice listening mode.
There are also cases in which no one of the LCO instruments is able to observe the target: in this case users are informed and the programme resumes the socket listening activities (Sect.~\ref{Telegram_not_obs_ex}).

\subsection{Submission}
\label{subsec:submission and getting products}

\begin{figure*}
\centering
\includegraphics[width=12cm]{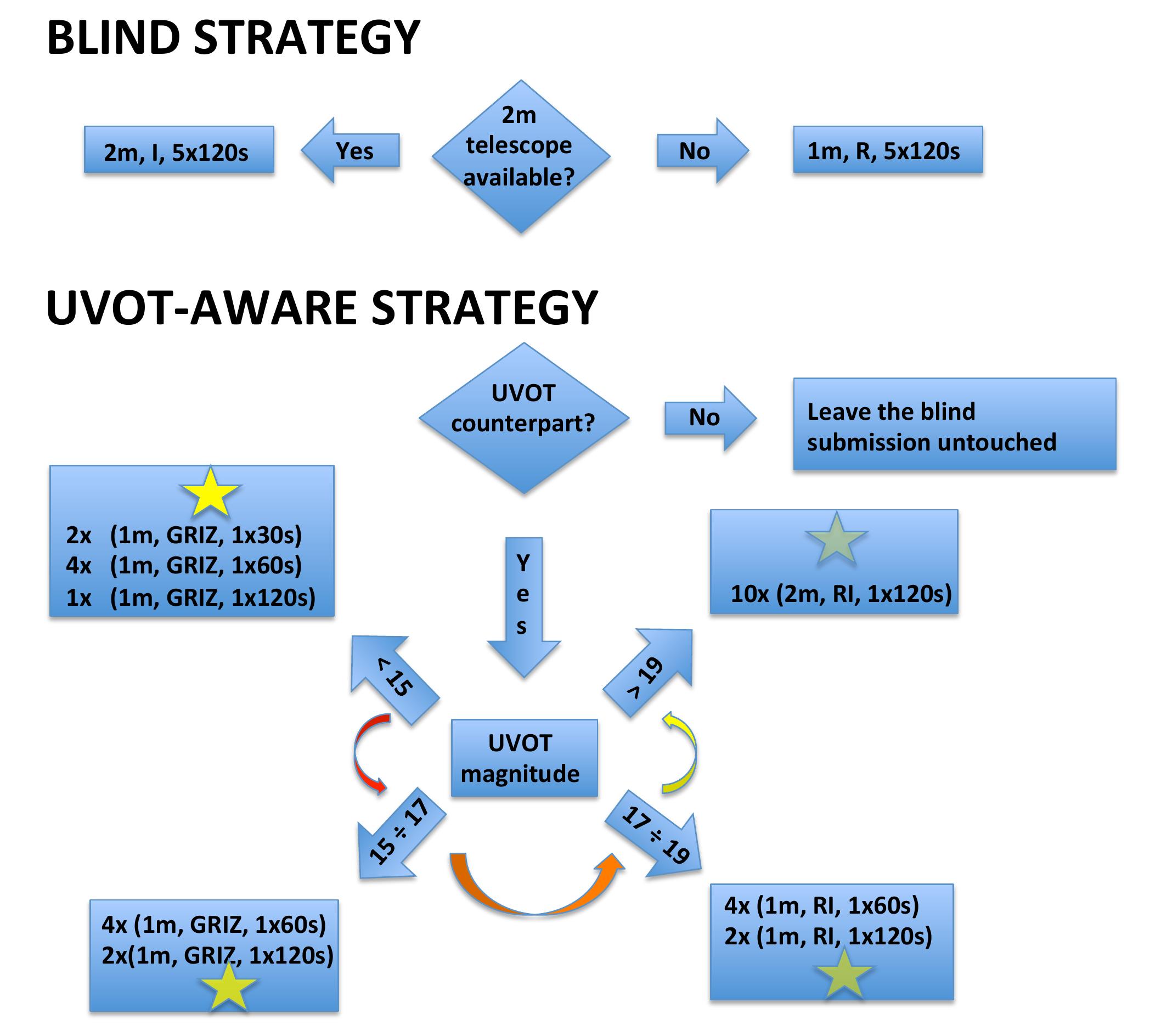}
\caption{
Sketch of the observation strategy, before and after UVOT information reception.
``Blind'' observation request describes the observing strategy when no information from UVOT is available (yet) and is based on the BAT estimated position. The only two-option choice is driven by the availability of 2m-class telescopes. The blind request is always submitted when the bursts is observable by at least one of the instruments we use. The blind step is the only one in the case of {\em INTEGRAL} burst Notices.
UVOT-aware observation request: observing strategy upon receiving a Notice of a UVOT detection within 15 minutes of the reception of a {\em Swift}-BAT Notice. If no UVOT counterpart detection has been detected, nothing more is done for that trigger; otherwise, the choice is driven by the estimated UVOT magnitude in the w filter, as reported in the corresponding GCN Notice.}
\label{fig:strategy}
\end{figure*}

The Neil Gehrels {\em Swift} Observatory working strategy relies on the complex and very efficient interplay between its three on-board instruments: the Burst Alert Telescope (BAT), the X-ray Telescope (XRT) and the UV/Optical Telescope (UVOT). When a signal is detected and identified by the large-field instrument (BAT), the satellite rapidly slews to point XRT and UVOT that observe at lower energies. The distributed {\em Swift} GCN Notices reflect this strategy. The first Notice is produced after the BAT trigger and contains the first rough estimation of the position (usually with an error radius of a few arcminutes), then, in the absence of observing constraints, XRT and UVOT are automatically pointed towards the source and more accurate information about position (with an error radius usually within a few arcsecs), flux and magnitude (for UVOT) is produced and distributed whenever credible candidates are found within the BAT error circle. The time lag between the BAT detection and XRT and UVOT pointing is quite variable, depending on a number of observational and technical constraints (see Section~\ref{subsec:UVOT statistics} for further details). A good strategy for a fast optical followup  should exploit UVOT Notices too, since the detection (or lack thereof) of an UVOT counterpart is a further piece of information that can be used to refine the data acquisition (see below).

The basic principle of our strategy for {\em Swift} is the prompt submission of a fast, preliminary observation request once the first burst notice from BAT has come along. Then, the GRB pipeline may send a second one, better tailored to the UVOT counterpart brightness, in the case of an optical detection by UVOT. Following this guideline, after the first submission, the pipeline waits for 15~minutes for a UVOT notice: in case of no optical detection, no additional requests are sent to the LCO network and the already-sent preliminary request becomes definitive. Differently, the GRB pipeline extracts from the UVOT Notice the refined position and the estimated optical magnitude in the White\footnote{\url{http://www.swift.ac.uk/analysis/uvot/filters.php}} filter.
A flow chart describing the different branches of possibilities is shown in Figure~\ref{fig:strategy}.

The strategy based on {\em INTEGRAL} Notices is simpler, given that the ``REFINED'' Notice is not followed by anything and there are no prompt re-pointings of different energy instruments. This is the reason why the strategy in this case is the same as the ``Blind'' one for {\em Swift} and no further steps are required.

The submission of an observation request, like almost all the other operations on the LCO network, can be done both via web interface\footnote{\url{https://observe.lco.global/create/}} and API script\footnote{\url{https://developers.lco.global/?python#observations}}.

\section{Statistical analysis}
\label{sec:statistics}

\subsection{Prompt UVOT information}
\label{subsec:UVOT statistics}
\begin{figure*}
\centering
\includegraphics[width=13cm]{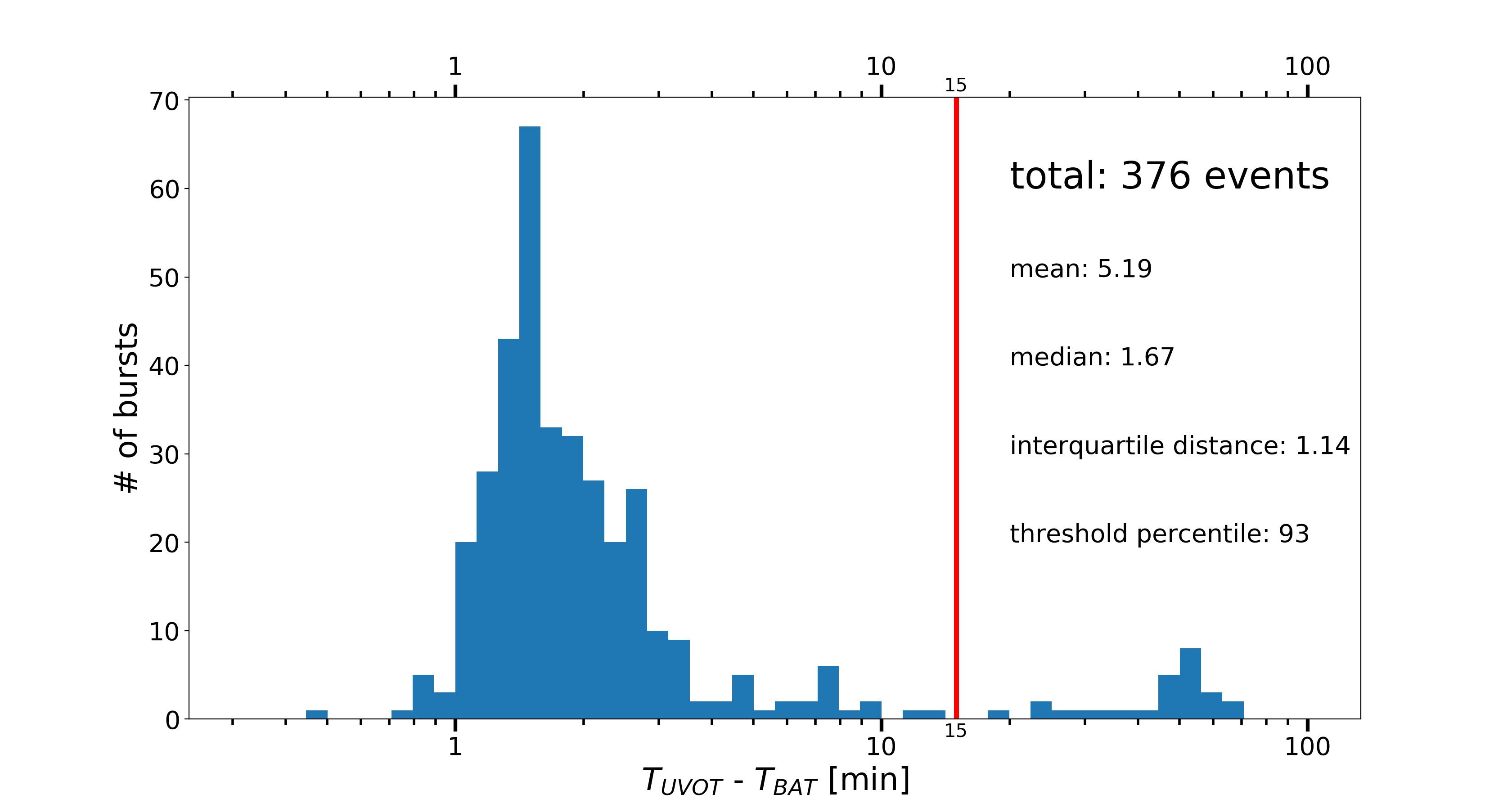}
\caption{Distribution of the time delay between the arrival of the {\em Swift}-UVOT Notice trigger and the {\em Swift}-BAT Notice trigger built on a dataset of 376 events. In $\sim90$\% of cases, this delay is $<15$~minutes. The distribution has been obtaining considering only events with $T_{UVOT}-T_{BAT}<100$ min, since larger deltas are usually connected with pathological cases.}
\label{fig:UVOT_stat}
\end{figure*}

\begin{figure*}
\centering
\includegraphics[width=13cm]{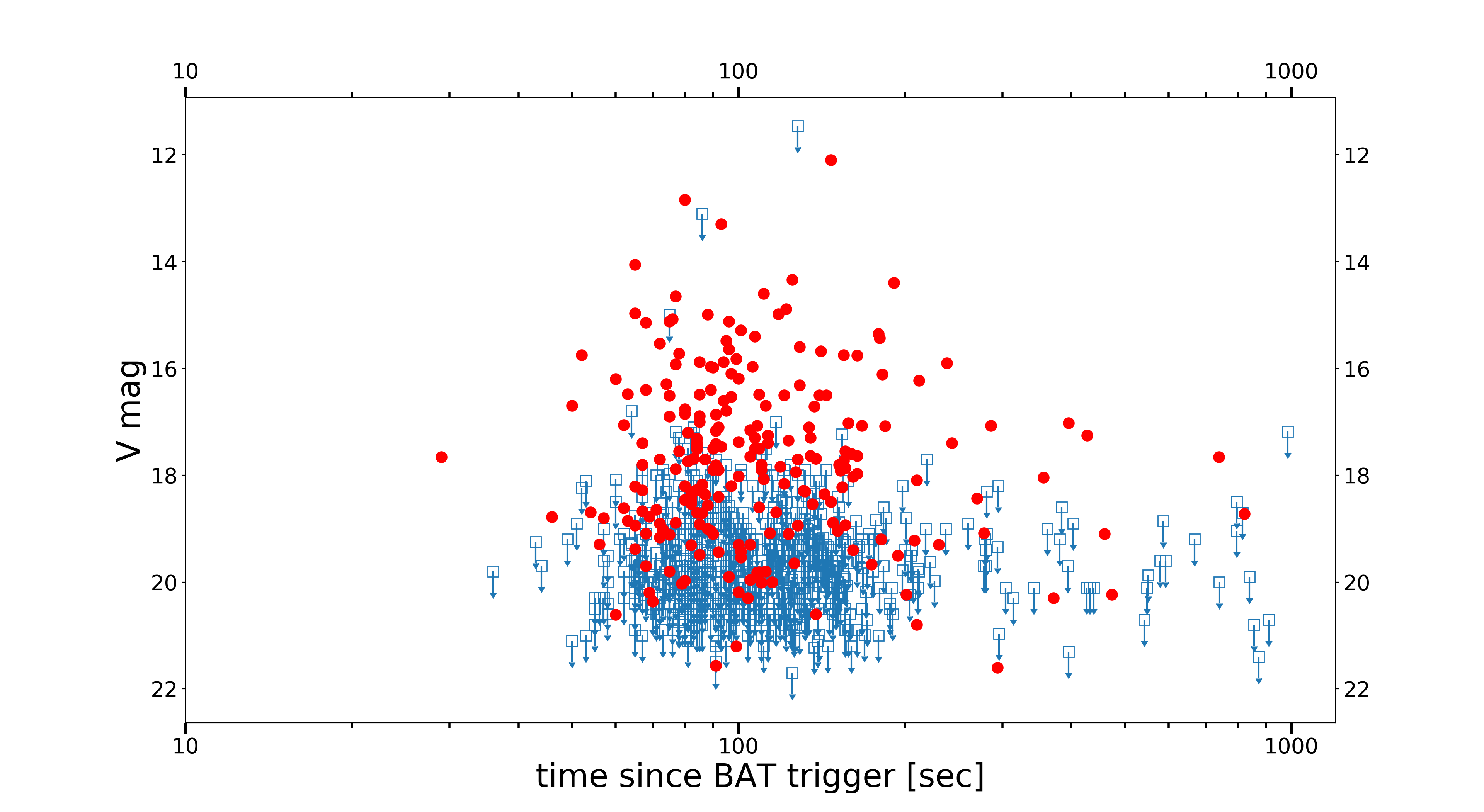}
\caption{Scatter plot of GRBs promptly observed by {\em Swift}-UVOT, showing V-mag vs. $T-T_{BAT}$. Red dots and blue squares show detections and upper limits, respectively. Unusually long delays ($T-T_{BAT}>1000$~s) have been omitted.}
\label{fig:UVOT_mag}
\end{figure*}

The existence of a time delay between the BAT trigger and the detection of a UVOT counterpart is something unavoidable, due to the {\em Swift} working strategy and affected by wide number of factors, such as the slewing time of the facility, and other observation constraints. 
This justifies the need for a waiting time between the ``Blind strategy'' and the ``UVOT-aware'' steps of our strategy (Figure~\ref{fig:strategy}).
Since a rapid and proper response to trigger notices is the first aim of the GRB pipeline, an optimal choice of this waiting time is fundamental: one has to find a trade-off between the need to wait for the UVOT information and the need to proceed quickly. To find the optimal duration of the waiting time, we derived the distribution of the delay times between the BAT position notice and the UVOT notice. We used the information available at the {\em Swift} web page\footnote{\url{https://swift.gsfc.nasa.gov/archive/grb_table/}}.
We included all {\em Swift} GRBs with a UVOT detection, that represent $\approx25\%$ of the total. We discarded all the bursts with a time delay $T_{\rm UVOT}-T_{\rm BAT}>100$~min that were due to delayed followups, most of which were probably due to observing constraints. The remaining dataset consists of 376 events and its $T_{\rm UVOT}-T_{\rm BAT}$ distribution and some related quantities are displayed in Fig.~\ref{fig:UVOT_stat}. The distribution is asymmetric, with a long tail extending to large time delays all the way up to the 100-minute threshold. Nonetheless, the vast majority of times clusters around a few minutes, as shown by the median ($\sim 1.7$~min) and interquartile span ($\sim1.1$~min). When no threshold on the BAT-UVOT delay is set, 93\% of GRBs with a UVOT counterpart have a delay shorter than 15~min. This makes us confident that the GRB pipeline can exploit UVOT information in more than 90\% cases. A clearer picture can be obtained looking at Fig. \ref{fig:UVOT_mag}, where we reported all the GRBs from start of {\em Swift} up to now in the V-mag/T-$T_{BAT}$ plane. Most detections occur in the first 200 s after the BAT trigger. The brightest events are mostly seen around $\sim100$~s and not earlier. This time roughly corresponds to the time of the peak of the distribution of optical afterglows. This is probably the result of a mere observation bias, since rare events are more likely to be found around the most sampled times.

\subsection{LCO response to ToO requests}
\label{subsec:LCO statistics}

\begin{figure}
\centering
\includegraphics[width=9cm]{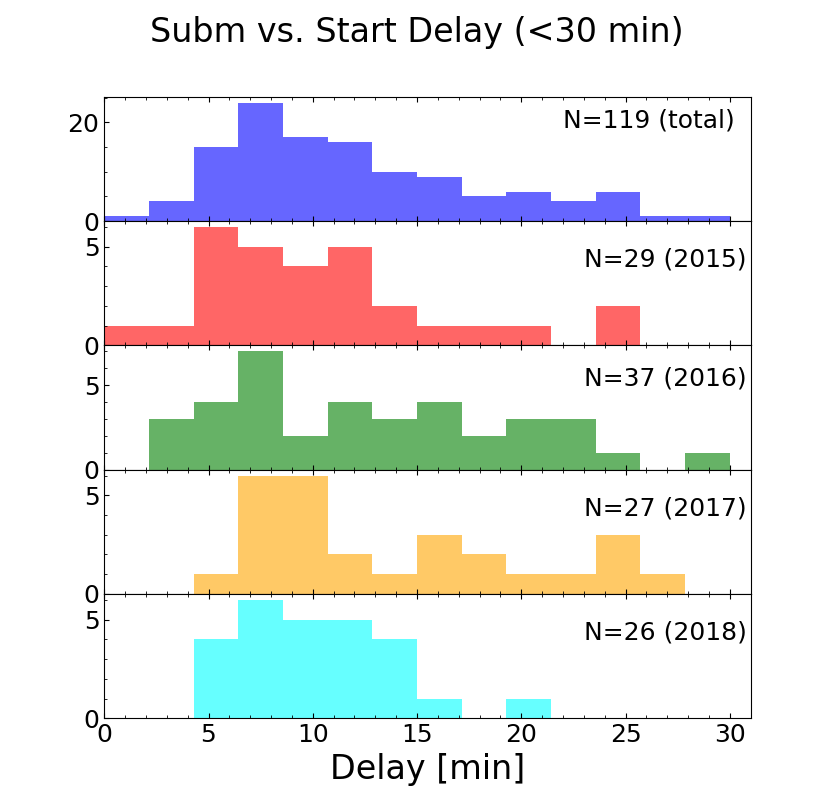}
\caption{Time delay between ToO request submission and observation start for our proposals on the LCO network from January 2015 up to now.}
\label{fig:LCO_stat_crg}
\end{figure}

\begin{figure}
\centering
\includegraphics[width=9cm]{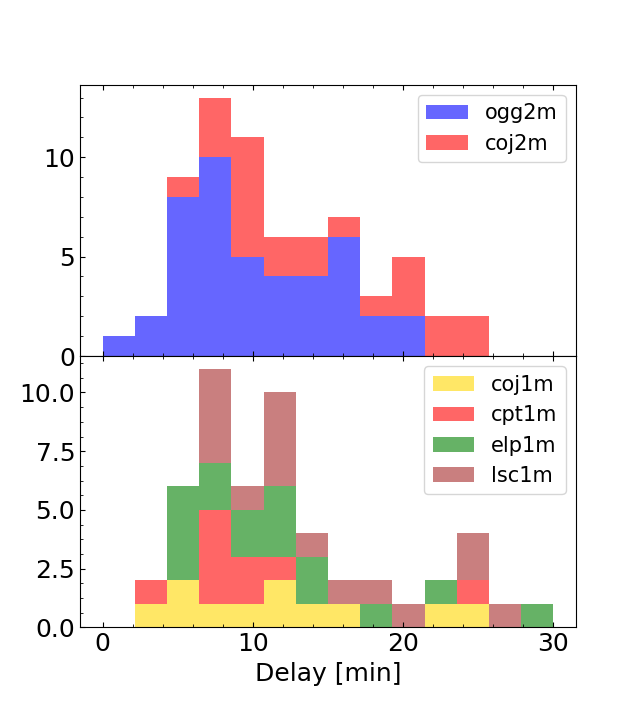}
\caption{Time delay between ToO request submission and observation start for our proposals on the LCO network from January 2015 up to now, for each individual telescope (color-coded): 2-m units ({\em top}), and 1-m units ({\em bottom}). A KS test comparing the distributions of ogg2m (Hawaii) vs. that of coj2m (Siding Springs) yields a $4.4$\% probability of being drawn from the same population, with ogg2m seemingly faster than coj2m on average.}
\label{fig:LCO_stat_crg_individ_units}
\end{figure}

\begin{figure*}
\centering
\includegraphics[width=13cm]{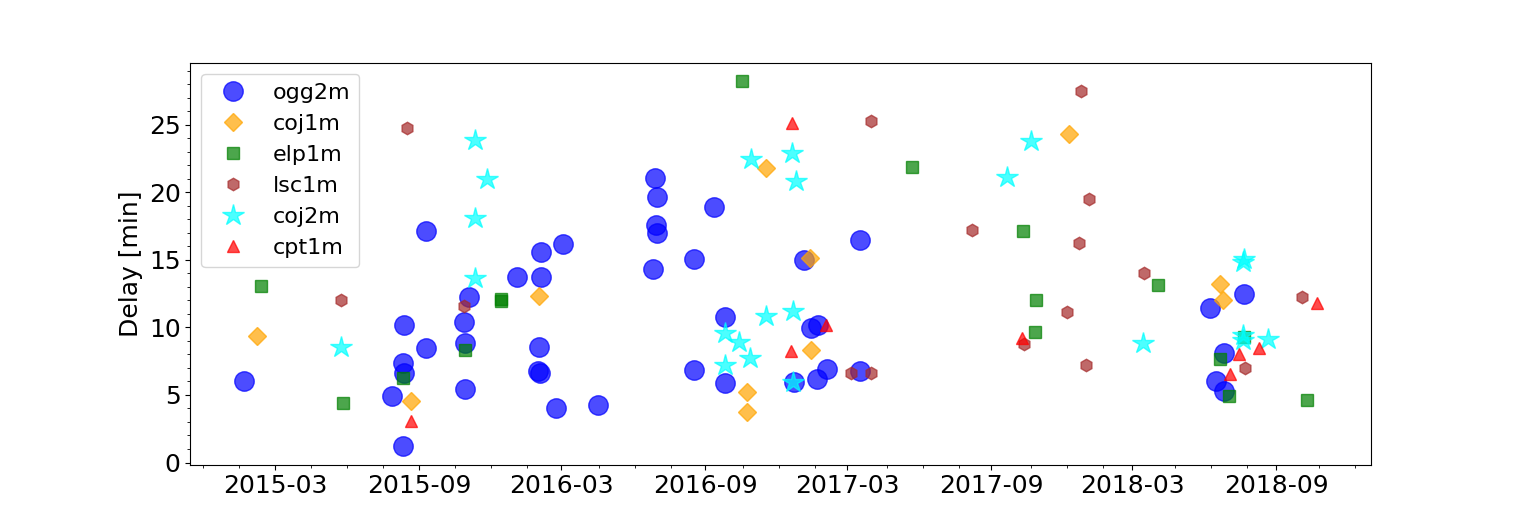}
\caption{Delay [minutes] between ToO submission time and effective start time as a function of epoch for our proposals on the LCO network, starting from January 2015 up to now. Markers uniquely indicate observatories and observing units accessible to our programme.}
\label{fig:LCO_scatter_crg}
\end{figure*}

To get a complete picture of the time delay from the reception of a GCN up to the frame acquisition, we analyse the time required by LCO to start acquiring data once an observation request is received. This delay depends mainly on the pressure of users on the network, meaning a large number of observation requests for the scheduler to manage. This number varies randomly and so does the time delay, too.
We built our time delay distributions considering observation requests from all our approved proposals on LCO (Section~\ref{tab:proposals}), covering the entire time span (Figures~\ref{fig:LCO_stat_crg} and \ref{fig:LCO_stat_crg_individ_units} and \ref{fig:LCO_scatter_crg}). The reason why the distribution has sligthly changed over the years is that the scheduler algorithms have continuously been adjusted in response to the various changes that occurred in the meanwhile. 

\section{Example of a successful case}
\label{sec:succ_case}

\begin{figure}
\centering
\includegraphics[width=9cm]{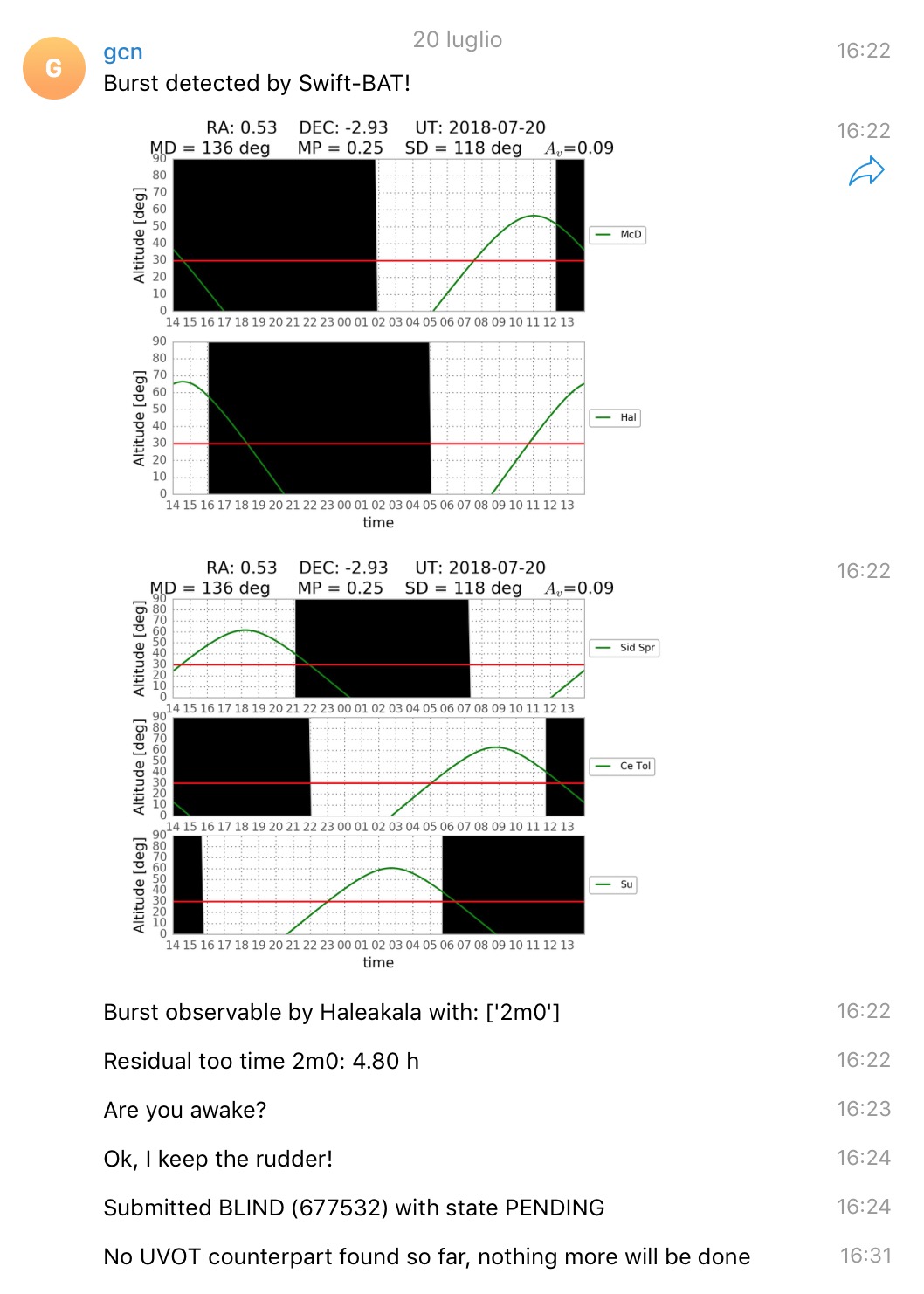}
\caption{Screenshot of the Telegram output for GRB 180720B. Note that times are in the CET frame, so GMT+2 at the moment of the burst. The pipeline indicated the burst as promptly visible from Haleakala, but a few minutes later the burst became observable from Siding Spring, that finally performed the observation as established by the scheduler.}
\label{fig:Telegram_succ}
\end{figure}

On July 20, 2018, at 14:21:44 UT {\em Swift}-BAT triggered GRB\,180720B ($\alpha$, $\delta$)= ($0.530^\circ,$ $-2.933^\circ$; J2000) and a GCN Notice \cite{Siegel18} was promptly distributed via both email and socket.
We received the socket message at 14:22:44 UT and the pipeline began operations\footnote{Please note that we used a slightly different procedure in this case, being in a first testing phase. The main differences where:
\begin{itemize}
    \item we used a 7-minutes waiting time for the UVOT counterpart;
    \item we submitted a slightly different ``blind'' observation request, composed by 5 exposure of 120s in the i' filter.
\end{itemize}} to convey burst information, wait for human intervention and then submit a ``blind'' observation request.
We were informed by the pipeline at 14:22:45 UT and the very first observation request was automatically submitted at 14:24:07 UT, after 1 minute of waiting for the human intervention. It is worth noting that the email Notice was received at 14:24, at the moment of our first submission.
The 2-m unit at Siding Spring Observatory started observing at 14:33:34 UT (710 s from the BAT trigger) and completed the sequence at 14:46:43 (1499 s from the BAT trigger) UT\footnote{The pipeline flagged the burst as promptly visible from Haleakala, but a few minutes later the burst became observable from Siding Spring, too. According to internal rules, the scheduler finally decided to convey our observation request towards Siding Spring, that finally performed the observation.} 
Data became available a few minutes after the end of the observation sequence and so we were able to promptly analyse the frame less than 30 minutes after the burst and to identify the optical afterglow. Our discovery was issued to the community about an hour after the explosion \cite{Martone18}.
Figure~\ref{fig:Telegram_succ} reports a screenshot for the Telegram set of notification for this burst we distributed to our users.

\section{Conclusions}
\label{sec:conc}

This work describes the automatic pipeline that we developed and used for fast followup of optical GRB counterparts with the LCO network. Designed to respond to triggers by {\em Swift} and {\em INTEGRAL} with the maximum rapidity allowed by LCO, proved to be successful in ensuring the monitoring of the first minutes of the  flux evolution of the GRB optical counterpart.
It is developed under Python environment, to ensure the highest cross-platform compatibility and to exploit the large amount of astronomy-oriented packages and utilities available.

The ability of fast reading the notices benefits from receiving alerts via socket, while  communicating the content to the users is handled via the Telegram platform. The advantages of this combination are the rapidity, reliability, and the absence of any cost notification delivering.

The submission step takes  advantage of the peculiarities of the LCO network, whose many eyes (spread all over the globe) are managed by a single, central scheduler in a full-robotic design. The scheduler is directly triggered by the pipeline observation requests, which for {\em Swift} bursts are tailored on the available UVOT information.

We described the example of GRB\,180720B, for which the pipeline enabled us to track the optical evolution of the burst a few minutes after the notice arrival. Specifically, observations were submitted 143 s and began 710 s after the BAT trigger  \cite{Martone18}.

Based on our experience in the Target-of-Opportunity observations with LCO network, we have carried out a statistical study of the response times of the whole system. The results are critical to understand the potential and current limitations of the network being used for any fast followup programme. Furthermore, they represent a starting point and a useful reference for future improvements and upgrades of the network. The ultimate goal of the pipeline is to optimise receipt and response to any programmes on the prompt followup of transient triggers, for which the study of the early phases is key to capture the encoded physics. While conceived for GRB followup, its portability and versatility make it a valuable reference for other fast transient programmes in the era of transient sky. The same configuration can be further exploited using notices from {\em Fermi} \cite{Meegan09} to trigger large-field-of-view ($>1^{\circ}\times1^{\circ}$) telescopes. The opportunities of following up rapid transients are due to dramatically increase in the next years, also thanks to the upcoming Large Synoptic Survey Telescope \cite{LSST09}.

\begin{acknowledgements}
We thank Leo Singer for its help and support in the developing of the code. Support for this work was provided by Universit\`a di Ferrara through grant FIR~2018 ``A Broad-band study of Cosmic Gamma-Ray Burst Prompt and Afterglow Emission" (PI Guidorzi).
\end{acknowledgements}

\bibliographystyle{spphys}
\bibliography{alles_grbs}

\end{document}